\documentclass[a4paper]{IEEEtran}
\usepackage[utf8x]{inputenc}
\usepackage{lipsum}
\usepackage{amssymb}
\usepackage{amsmath}
\usepackage{cite}   
\usepackage{slashbox,booktabs}
\usepackage{multirow}
\usepackage{graphicx}
\usepackage{upgreek}

\newcommand{\xb}{{\boldsymbol x}}
\newcommand{\yb}{{\boldsymbol y}}

\title{Improving the Spectral Efficiency of Nonlinear Satellite Systems through Time-Frequency Packing and Advanced Processing}
\author{Amina Piemontese,~\IEEEmembership{Member,~IEEE}, Andrea Modenini,~\IEEEmembership{Student Member,~IEEE},  Giulio Colavolpe,~\IEEEmembership{Senior Member,~IEEE}, and Nader Alagha,~\IEEEmembership{Member,~IEEE}
\thanks{G. Colavolpe, A. Modenini, and A. Piemontese are with are with Consorzio Nazionale Interuniversitario per le Telecomunicazioni (CNIT) and Universit\`a di Parma, Dipartimento di Ingegneria dell'Informazione, 
Viale G.~P.~Usberti, 181A, I-43124 Parma, Italy, e-mail: giulio@unipr.it, modenini@tlc.unipr.it, amina.piemontese@unipr.it. N. Alagha 
is with the European Space Agency (ESA-ESTEC), Noordwijk, The Netherlands}
\thanks{This work is funded by the European Space Agency, ESA-ESTEC, Noordwijk, The Netherlands, under contract no. 4000102300.}
\thanks{The paper was presented in part at the IEEE Intern. Conf. Commun. (ICC'12), Ottawa, Canada, June 2012.}
}

\begin{document}

\maketitle

\begin{abstract}
We consider realistic satellite communications systems for broadband and broadcasting applications, based on frequency-division-multiplexed linear modulations, where spectral efficiency is one of the main figures of merit. For these systems, we investigate their ultimate performance limits by using a framework to compute the spectral efficiency when suboptimal receivers are adopted and evaluating the performance improvements that can be obtained through the adoption of the time-frequency packing technique. Our analysis reveals that introducing controlled interference can significantly increase the efficiency of these systems. Moreover, if a receiver which is able to account for the interference and the nonlinear impairments is adopted, rather than a classical predistorter at the transmitter coupled with a  simpler receiver, the benefits in terms of spectral efficiency can be even larger. Finally, we consider practical coded schemes and show the potential advantages of the optimized signaling formats when 
combined with iterative detection/decoding.
\end{abstract}

\section{Introduction}
In satellite systems, orthogonal signaling is often adopted to avoid intersymbol interference (ISI), at least in the absence of nonlinear distortions. For example, 
in the 2nd-generation satellite digital video broadcasting (DVB­-S2) standard~\cite{DVB-S2-TR}, 
a conventional square-root raised-cosine (RRC) pulse shaping filter is specified at the transmitter. 
For an additive white Gaussian noise (AWGN) channel and in the absence of other impairments, the use of a RRC filter at the receiver 
and proper sampling ensure that optimal detection can be performed on a symbol-by-symbol basis. 
On the other hand, it is known that, when finite-order constellations are considered  
(e.g., phase-shift keying (PSK)), the spectral efficiency (SE) of the communication system, defined here as the information rate normalized to the spectral bandwidth assigned to the channel and the time spent to transmit a symbol, can be improved by relaxing the orthogonality condition.
\textit{Faster-than-Nyquist} signaling (FTN, see~\cite{Ma75c,MaLa88,LiGe03,RuAn05}) is a well known technique 
consisting of reducing the time spacing between two adjacent pulses well below that ensuring the Nyquist condition, thus introducing ISI. 
If the receiver is able to cope with the interference, the efficiency of the communication system will be increased.
In the original papers on FTN  signaling~\cite{Ma75c,MaLa88,LiGe03,RuAn05}, this optimal time spacing is obtained as the smallest value giving 
no reduction of the minimum Euclidean distance with respect to the Nyquist case. 
This ensures that, asymptotically, the ISI-free bit-error rate (BER) performance is reached, at least when the optimal detector is adopted. The i.u.d. capacity 
or information rate (IR), i.e., the average mutual information when the channel inputs are 
independent and uniformly distributed (i.u.d.) random variables, is then computed, still assuming the adoption of the optimal detector~\cite{RuAn06,RuAn07}. 
However, the complexity of this optimal detector easily becomes 
unmanageable, and no hints are provided on how to perform the optimization in the more practical scenario where a reduced-complexity receiver is employed. 

In~\cite{BaFeCo09b}, for the AWGN channel, a different approach for improving the SE has been considered.  
The approach relies on both time packing of adjacent symbols and reducing the carrier spacing of the adjacent channels when applicable (multi­carrier transmission), 
thus introducing
also interchannel interference (ICI). 
In~\cite{BaFeCo09b} it is assumed that a symbol-by-symbol detector 
working on the samples at the matched filter output is adopted at the receiver side, and the corresponding IR is computed, by also optimizing time 
and frequency spacings to maximize the  achievable SE. Hence, rather than the 
minimum distance, and thus the BER, the SE is the performance figure of merit. In addition, a low-complexity memoryless receiver is considered rather than the optimal 
detector employed in~\cite{Ma75c,MaLa88,LiGe03,RuAn05,RuAn06,RuAn07}. 
More complex detection algorithms have been also considered in~\cite{MoCoAl12} for the case of linear channels. 

In this work, we apply the time-frequency packing (TF packing) technique to nonlinear satellite channels. In particular, we design highly efficient schemes by choosing 
the time and frequency spacings which give the maximum value of SE. 
We assume a realistic satellite channel where nonlinear distortions originate from the presence of a high-power amplifier (HPA).
The considered system is also affected by ISI, due to the presence of input and output multiplexing (IMUX and OMUX) filters placed before and after the HPA and intentionally introduced by the adoption of the time packing technique as well. Although ICI is also present due to frequency packing,
we limit our investigation to systems in single-carrier-per transponder operation (i.e., each transponder is devoted to the amplification of the signal coming from only one user) where a single-user receiver is employed, and consider two different approaches 
to detection for nonlinear channels,
namely the use of a detector taking into account the nonlinear effects and a more 
traditional scheme based on predistortion and memoryless detection.
In the case of predistortion, we consider the dynamic data predistortion technique described in~\cite{KaSa91,CaDeGi04}, whereas in the case of advanced detection we 
consider a receiver based on an approximate signal model described in~\cite{CoPi12}. In this latter case, we also employ the channel shortening technique
(CS)~\cite{CoMoRu12}, recently proposed for nonlinear satellite channels, to limit the complexity of the detection algorithm. 
It should be noted that we apply the TF
packing technique to nonlinear satellite channels for which, usually,  even by using RRC pulses there is still ISI at the receiver.

The proposed TF packing technique promises to provide increased SEs at least for low-order modulation formats. In fact, when dense constellations with shaping are employed, 
we fall in a scenario similar to that of the Gaussian channel with Gaussian inputs for which orthogonal signaling with no excess bandwidth (rectangular shaping pulses) is optimal (although this is mainly true for the linear channel and not in the presence of a nonlinear HPA, 
since shaping increases the peak-to-average power ratio). Improving the achievable SE without increasing the constellation order can be considerably convenient since it is well known that low-order constellations are more robust to channel impairments such as time-varying phase noise and non-linearities. It is expected that the use of low-order modulation in conjunction with TF packing provides similar advantages in terms of robustness against channel impairments.

The proposed approach to improve the SE is very general and the  case of satellite systems for broadband and broadcasting applications must be thus considered just as an example to illustrate the benefits that can be obtained through the application of the TF packing paradigm coupled with advanced receiver processing.

The remainder of this paper is organized as follows. In Section~\ref{s:system_model}, we introduce the system model. 
The framework that we use to evaluate the SE of satellite systems is detailed in Section~\ref{s:spectral efficiency}, whereas different approaches to the detection 
for the considered channel are described in Section~\ref{s:auxiliary channels}. Numerical results are reported in Section~\ref{s:results}, where we show how the 
proposed technique can improve the SE of DVB-S2 systems. Finally, conclusions are drawn in Section~\ref{s:conclusions}.

\section{System Model}~\label{s:system_model}
We consider the forward link of a transparent satellite system, where synchronous users employ the same linear modulation format, shaping pulse $p(t)$, and symbol interval 
(or time spacing) $T$, and access the channel according to a frequency division multiplexing scheme. 
The transmitted signal in the uplink can be expressed as 
\begin{equation}\label{e:transmitted signal}
x(t)=\sum_\ell \sum_k x^{(\ell)}_{k} p(t-kT)e^{j2\pi \ell F_u t}\, ,
\end{equation}
where $x^{(\ell)}_{k}$ is the symbol transmitted by user $\ell$ during the $k$th symbol interval, and $F_u$ is the frequency spacing between adjacent channels.\footnote{In this scenario, we will use the terms,
``channels'', ``users'', and ``carriers'' interchangeably.}
The transmitted symbols belong to a given zero-mean $M$-ary complex constellation.
In \mbox{DVB-S2} standard the base pulse $p(t)$ is an RRC-shaped pulse with roll-off factor $\alpha$ (equal to 0.2, 0.3, or 0.35 depending on the service requirements). Notice that, in order to leave out border effects, the summations 
in~\eqref{e:transmitted signal} extend from $-\infty$
to $+\infty$, namely an infinite number of time epochs and carriers
are considered. 

\begin{figure}[t]
\centering{}\includegraphics[width=88mm]{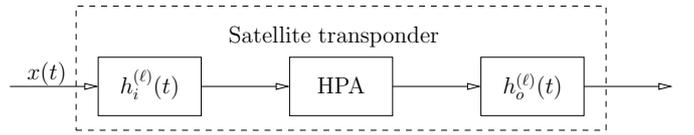} 
\caption{Satellite transponder for user $\ell$.{\label{f:trasponder}}}
\end{figure}

As commonly assumed for broadband and broadcasting systems, on the feeder uplink (between the gateway and the satellite) the impact of thermal noise can be neglected due to a high transmit signal strength. Hence, in our analysis, we have considered a noiseless feeder uplink.
Although the TF packing can be applied to other and more general scenarios,  we consider here a single-carrier per transponder scenario, where different carriers undergo independent amplification by different transponders on board of satellite, each of which works with 
a single carrier occupying its entire bandwidth.
In this case, the on-board power amplifier can operate closer to saturation and hence improve its efficiency. 
The transponder model for user $\ell$, shown in Fig.~\ref{f:trasponder}, is composed of an IMUX filter $h_i^{(\ell)}(t)$ which selects the $\ell$th carrier, an HPA, 
and an OMUX filter  $h_o^{(\ell)}(t)$ 
which reduces the out-of-band power due to the spectral regrowth after nonlinear amplification~\cite{DVB-S2-TR}.
The HPA is a nonlinear memoryless device defined through its AM/AM and AM/PM characteristics, describing the amplitude and phase distortions caused on the signal at 
its input. 

The outputs of different transponders are multiplexed again in the downlink to form the signal $s(t)$, and we assume that the adjacent users have a frequency 
separation of $F_d$, usually equal to that in the uplink. 
The useful signal at the user terminal is still the sum of independent contributions, one for each transponder (although these contributions are no more, rigorously, 
linearly modulated due to the nonlinear transformation of the on-board HPA).
The received signal is also corrupted by the downlink AWGN, whose low-pass equivalent  $w(t)$ 
has power spectral density (PSD) $2N_{0}$. The low-pass equivalent of the received signal has thus expression
\begin{equation}\nonumber
 r(t)=s(t)+w(t) \, .
\end{equation}

We evaluate the ultimate performance limits of this communication system when single-user detection is employed at the receiver side. 
The proposed technique consists of allowing interference in time and/or frequency by reducing the values of $T$, $F_u$, and $F_d$, (partially) coping with it at the 
receiver, in order to increase the SE. 
In other words, $T$, $F_u$, and $F_d$ are chosen as the values that give the maximum value of the SE. 
These values depend on the employed detector---the larger the interference that the receiver can cope with, the larger the SE and the lower the values of time and 
frequency spacings.
Notice that, since we are considering single-user detection, the receiver is not able to deal with the interference due to the overlap of different channels.
In this case, the optimization of the frequency spacings is actually an optimization of the frequency guard bands generally introduced in satellite systems to avoid 
the nonlinear cross-talk, since a single-user receiver can tolerate only a very small amount of ICI.

The considered nonlinear satellite channel reduces, as a particular case, to the linear channel, provided that the HPA is driven far from saturation. Hence, all the considerations in this paper can be straightforwardly extended to the linear channel case. A few results can be found in~\cite{MoCoAl12}.

\section{Optimization of the Spectral Efficiency}\label{s:spectral efficiency}
We describe the framework used to evaluate the ultimate performance limits of the considered satellite system and to perform the optimization of the time and 
frequency spacings. 
To simplify the analysis, we will assume $F_u=F_d=F$. 
We perform this investigation by constraining the complexity of the employed receiver. In particular, as mentioned, we assume that a single-user detector is used. 
For this reason, without loss of generality we only consider the detection of symbols $\xb^{(0)}=\{x^{(0)}_k\}$ of user with $\ell=0$.\footnote{Assuming a system with 
an infinite number of users, the results do not depend on a specific user.} In addition, we also consider low-complexity receivers taking into account only a portion 
of the actual channel memory. Under these constraints, we compute the IR, i.e., the average mutual information when the channel inputs are 
i.u.d. random variables belonging to a given constellation. Provided that a proper \textit{auxiliary channel} can be defined 
for which the adopted low-complexity receiver is optimal, the computed IR represents an achievable lower bound of the IR of the actual channel, according to 
mismatched detection~\cite{MeKaLaSh94}.

Denoting by $\yb^{(0)}$  a set of sufficient statistics for the detection of $\xb^{(0)}$, the achievable IR, measured in bit per channel use, can be obtained as
\begin{equation}\label{e:IR}
I(\xb^{(0)} ; \yb^{(0)}) =\lim_{K\rightarrow \infty}\frac{1}{K} E
\Bigg\{\log \frac{p(\yb^{(0)} \arrowvert \xb^{(0)} )}{p(\yb^{(0)})}  \Bigg\}\, ,
\end{equation}
where $K$ is the number of transmitted symbols. The probability density functions $p(\yb^{(0)} \arrowvert \xb^{(0)})$
and $p(\yb^{(0)})$ are computed by using the optimal maximum-a-posteriori (MAP) symbol detector for the auxiliary channel, while the expectation in~(\ref{e:IR}) 
is with respect to the input and output sequences generated according to the actual channel model~\cite{ArLoVoKaZe06}. In the next section, we will discuss two 
different low-complexity detectors for nonlinear satellite channels and we will define the corresponding auxiliary channels.

We can define the user's bandwidth as the frequency separation $F$ between two adjacent carriers. The achievable SE is thus
\begin{equation}
\eta=\frac{1}{FT} I(\xb^{(0)} ; \yb^{(0)})  \quad[
\textrm{b/s}/\textrm{Hz} ] . \label{e:eta}
\end{equation}
The aim of the proposed technique is to find the values of $F$ and $T$ providing, for each value of the signal-to-noise ratio (SNR), the maximum value of SE achievable by that particular receiver, optimal for the considered specific auxiliary channel. Namely, we compute
\begin{equation}
\eta_\text{M}=\max_{F,T>0} \eta(F,T) \, . \label{e:eta_M}
\end{equation}
Typically, the dependence on the SNR value is not critical, in the sense that we can identify two or at most three SNR regions for which the 
optimal spacings practically have the same value.

For fair comparisons in terms of SE, we need a proper definition of the SNR. 
We define the SNR as the ratio $P/N$ between the transmit power when the HPAs are driven at saturation and the noise power (in the considered bandwidth). 
Denoting by $U$ the number of users and by $W$ the signal bandwidth, $P/N$ can as written as
\begin{equation}
  { P \over N } = \lim_{U \rightarrow \infty} { U P_{\mathrm{sat}} \over N_0 ( (U-1) F + W) } = { P_{\mathrm{sat}} \over N_0F } \,, \label{e:SNR}
\end{equation}
where $P_{\mathrm{sat}}$ is the peak power at the output of an  HPA in response to a continuous wave input, denoted as the amplifier saturation power. $P_{\mathrm{sat}}$ is independent of the bandwidth $W$. The SNR definition as given in (\ref{e:SNR}) is independent of the transmit waveform and its parameters. This provides a common measure to compare the performance of different solutions in a fair manner. 
The output back-off for each waveform (or modulation scheme) is defined as the power ratio (in dBs) between the unmodulated carrier at
saturation and the modulated carrier after the OMUX. 
We point out that equations \mbox{(\ref{e:eta})-(\ref{e:SNR})}, although derived for nonlinear satellite channels, can be used for linear AWGN channels also, by replacing $P_\mathrm{sat}$ in~(\ref{e:SNR}) with the user's transmitted power.

Without loss of generality, $T$ and $F$ in (\ref{e:eta})-(\ref{e:SNR}) can be normalized to some reference values $T_B$ and $F_B$.
We will denote $\nu=F/F_B$ and $\tau=T/T_B$. In the numerical results, we will choose $T_B$ and $F_B$ as the symbol time and the frequency spacing adopted in the DVB-S2 standard, which is considered as a benchmark scenario. 

\section{Auxiliary Channel Models}\label{s:auxiliary channels}

The system model described in Section~\ref{s:system_model} is representative of the considered scenario and has been employed in the information-theoretic analysis and in the simulations results. In this section, we describe the employed auxiliary channel models and the corresponding optimal MAP symbol detectors. As explained in Section~\ref{s:spectral efficiency}, they are used to compute two lower bounds on the SE for the considered channel~\cite{ArLoVoKaZe06}. Since these lower bounds are achievable by those receivers, we will say that the computed lower bounds are the SE values of the considered channel when those receivers are employed.

\subsection{Memoryless model and predistortion at the transmitter}\label{ss:predistortion}
When the HPA AM/AM and AM/PM characteristics are properly estimated and feeded back to the transmitter, the sequence of symbols $\{x^{(\ell)}_k\}$ can be properly 
predistorted to form the sequence  $\{x'^{(\ell)}_k\}$ that is transmitted instead, in order to compensate for the effect of the non-linearity and possibly to reduce the ISI. 
Here we consider the dynamic data predistortion technique described in~\cite{KaSa91,CaDeGi04} and also suggested for the application in DVB-S2 systems~\cite{DVB-S2-TR}, where the symbol $x'^{(\ell)}_k$ transmitted by user $\ell$ at time $k$ is a function of a sequence of $2L_p+1$ input symbols, i.e.,  $x'_k=f(x_{k-L_p},\ldots,x_k,\ldots,x_{k+L_p})$.
The mapping  $f$ at the transmitter is implemented through a look-up table (LUT), which is computed through an iterative
procedure performed off-line and described in~\cite{KaSa91,CaDeGi04}.  This procedure searches the best trade-off
between the interference reduction and the increase of the OBO.
The complexity at the transmitter depends on the number of symbols accounted for through the parameter $L_p$. 
The transmitted signal is thus
\begin{equation}\nonumber
 x(t)=\sum_\ell\sum_k x'^{(\ell)}_kp(t-kT)e^{j2\pi\ell Ft}
\end{equation}
whereas, at the receiver, a simple single-user memoryless channel is assumed corresponding to the auxiliary channel (for user with $\ell=0$)
\begin{equation}\label{e:auxiliary_channel_pred}
  y^{(0)}_k=x^{(0)}_k+n_k
\end{equation}
where $n_k$ is a zero-mean circularly symmetric white Gaussian
process with PSD $2(N_0+N_I)$, $N_I$ being a design parameter which
can be optimized through computer simulations---an increase of the assumed noise variance can improve 
the computed achievable lower bound on the spectral efficiency~\cite{BaFeCo09b}.

\subsection{Model with memory and advanced detection}\label{ss:advanced_detection}
A valid alternative to nonlinear compensation techniques at the transmitter relies upon the adoption of advanced detectors which can manage the nonlinear 
distortions and the ISI. 
In this work, we consider detection based on an approximate signal model described in~\cite{CoPi12}, which comes from a simplified 
Volterra series expansion of the nonlinear channel. 
In the following, we will consider PSK and  amplitude/phase shift keying (APSK) modulations 
typically employed in satellite systems~\cite{DVB-S2-TR}. To limit the receiver complexity with a limited performance degradation, we also apply a CS technique~\cite{FaMa73}.
In fact, when the memory of the channel is too large to be taken into account by a full complexity detector, an excellent performance can be achieved 
by properly filtering the received signal before adopting a reduced-state detector~\cite{FaMa73}. 
A very effective CS technique for general linear channels is described in~\cite{RuPr12}, while its extension to nonlinear satellite systems is reported 
in~\cite{CoMoRu12}.

The approximate model (employed for receiver design purposes only) of
the contribution $s^{(0)}(t)$ of user $\ell =0$ to the received useful signal $s(t)$, can be based on the following $v$th-order (with $v$
being any odd integer) \textit{simplified} Volterra-series expansion~\cite{CoPi12}
\begin{equation}
 s^{(0)}(t)\simeq\sum_{n} \sum_{i=0}^{N_V-1} \!x^{(0)}_n \left[ |x_n^{(0)}|^{2i} h^{(2i+1)}\!(t-nT) \right], \label{e:volterra_apprx} 
\end{equation}
where $N_V=(v+1)/2$, and $h^{(2i+1)}\!(t)$ are complex waveforms given by a linear combinations of the the original $N_V$ Volterra kernels. This 
simplified Volterra-series expansion is obtained from the classical one by neglecting some 
selected terms. For further details, the reader can refer to~\cite{CoPi12}. We are looking for an auxiliary channel and the corresponding optimal MAP symbol detector. As mentioned, only single-user receivers are considered here. Hence, we will assume that, apart from AWGN, only user $\ell=0$ is present and that the simplified model (\ref{e:volterra_apprx}) holds. 
Optimal single-user MAP symbol detection for this case can be performed through a bank of filters followed by a conventional BCJR detector~\cite{BaCoJeRa74} with 
proper branch metrics and working 
on a trellis whose number of states exponentially depends on the channel memory~\cite{CoPi12}.

In the case of PSK modulations, being $|x_i|^2=1$, the signal model (\ref{e:volterra_apprx}) simplifies to a linear modulation with shaping pulse
$\bar{h}(t)=\sum_{i=0}^{N_V-1} h^{(2i+1)}(t)$~\cite{CoPi12}. An approximate (being the model in (\ref{e:volterra_apprx}) an approximation) set of sufficient statistics $\yb^{(0)}$ can thus be obtained by sampling the output of a filter 
matched to $\bar{h}(t)$. Under the assumption that only user with $\ell=0$ is present, the $k$th element of $\yb^{(0)}$ is
\begin{equation}\nonumber
  y_k^{(0)} = \sum_i x^{(0)}_{k-i} g_i + \eta_k\, ,
\end{equation}
where
$$
g_i = \int \bar{h}(t)\bar{h}^*(t-iT) \mathrm{d}t$$ 
and $\eta_k$ is a Gaussian process with $\mathrm{E}\{ \eta_{k+i}\eta_k^* \}=2N_0g_i$.
Vector $\yb^{(0)}$ can be written as
\begin{equation}
 \boldsymbol{y}^{(0)} = \mathbf{G}\boldsymbol{x}^{(0)}+ \boldsymbol{\eta} \label{e:sufficient_statistics}
\end{equation}
where $\mathbf{G}$ is a Toeplitz matrix obtained from the sequence \{$g_i$\} whereas $\boldsymbol{\eta}$ is a vector collecting the noise samples.
According to the CS approach, the considered auxiliary channel is based on the following channel law~\cite{RuPr12}
\begin{equation}
  p(\yb^{(0)}|\xb^{(0)}) \propto \exp\left( 2\mathcal{R}({\yb^{(0)}}^\dagger \mathbf{H}^r \xb^{(0)}) - {\xb^{(0)}}^\dagger\mathbf{G}^r \xb^{(0)} \right) \, , 
\label{e:mis_channel_law}
\end{equation}
where $\mathbf{H}^r$ and $\mathbf{G}^r$ are Toeplitz matrices obtained from proper sequences
\{$h_i^r$\} and \{$g_i^r$\}, and are known as \textit{channel shortener} and \textit{target response}, respectively~\cite{RuPr12}. 
Matrix $\mathbf{H}^r$ represents a linear filtering  of the sufficient statistics~(\ref{e:sufficient_statistics}), and $\mathbf{G}^r$ 
is the ISI to be set at the detector (different from the actual ISI)~\cite{RuPr12}.
In~(\ref{e:mis_channel_law}), the noise variance has been absorbed into the two matrices. In order to reduce the complexity, we constrain the target response used at the receiver to
\begin{equation}
g^r_{i}={ 0} ~~~ |i|>L_{r} \label{e:G_constraint_psk}\, ,
\end{equation}
which implies that the memory of the detector is $L_{r}$ instead of the true memory of the channel. 
The CS technique finds a closed form of the optimal \{$h^r_i$\} and \{$g^r_i$\} which maximize the achievable IR~(\ref{e:IR}).
We point out that if the memory $L_r$ is larger than or equal to the actual channel memory the trivial solution is
$\mathbf{H}^r=\mathbf{I}/2N_0$ and $\mathbf{G}^r=\mathbf{G}/2N_0$, where $\mathbf{I}$ is the identity matrix.

For APSK modulations, an approximate set of sufficient statistics can be obtained through a bank of filters matched
to the pulses $h^{(2i+1)}(t)$, \mbox{$i=0,...,N_V-1$}.
The CS technique has been recently extended to this case in~\cite{CoMoRu12}, and leads to an auxiliary channel whose law  has the same form of~(\ref{e:mis_channel_law}), where the channel shortener 
and the target response are block matrices (see ~\cite{CoMoRu12} for details). 

\begin{figure}
\centering{}  \includegraphics[width=1.\columnwidth]{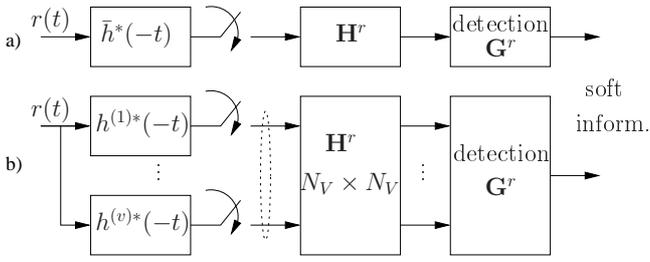}
  \caption{Block diagram of the considered suboptimal receivers for a) PSK and b) APSK modulations.}\label{f:receiver_cs}
\end{figure}

The resulting receivers for PSK and APSK modulations are shown in Fig.~\ref{f:receiver_cs}. They are optimal for the considered auxiliary channel models.
Interestingly, when $L_r=0$ the optimal channel shortener becomes the minimum mean square error (MMSE) feedforward equalizer of \cite{BeBi83}, applied to the signal 
model~(\ref{e:volterra_apprx}).

\section{Numerical Results}\label{s:results}
\subsection{Spectral efficiency}
Considering the DVB-S2 system as a benchmark scenario, we now
show the improvement, in terms of SE, that can be obtained by adopting the TF packing technique joint with an advanced processing at the receiver. Let us consider a typical DVB-S2 scenario where, at the transmitter, the shaping 
pulse $p(t)$ has a RRC spectrum, whereas the IMUX and OMUX filters and the nonlinear characteristics of the HPA are those reported in~\cite[Figs.~H.12 and H.13]{DVB-S2-TR}. 
The standard considers the following modulation formats: QPSK, 8PSK, 16APSK, and 32APSK. 
To combat ISI and nonlinear distortions, a data predistorter is employed at the transmitter whereas at 
the receiver a symbol-by-symbol detector is assumed.
Here, we consider the predistorter described in Section~\ref{ss:predistortion}, with $L_p=2$ for QPSK, 8PSK 16APSK, and $L_p=1$ for 32APSK.
The SE results have been obtained by computing the IR in~(\ref{e:IR}) by means of the Monte Carlo method described in~\cite{ArLoVoKaZe06}. For each case, we also performed a coarse optimization of the noise variance to be set at receiver~\cite{BaFeCo09b}, 
and of the amplifier operation point through the OBO. Unless otherwise specified, the roll-off factor of the RRC pulses is $\alpha=0.2$, which is the lowest value considered in the standard. 

\begin{figure}
\centering{} \includegraphics[width=92mm]{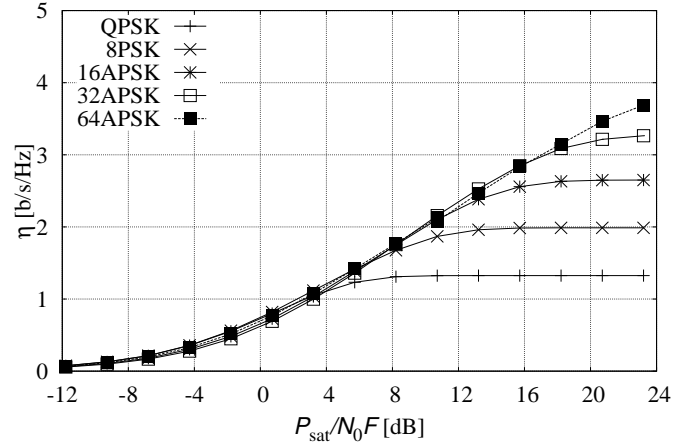}
 \caption{Spectral efficiency of DVB-S2 modulations with roll-off 0.2, data predistortion, and memoryless detection. Comparison with a constellation of increased cardinality (64APSK).}\label{f:ASE_DVB}
\end{figure}

We first consider the achievable SE of our benchmark scenario, and in Fig.~\ref{f:ASE_DVB} we report $\eta$ as a function of $P_{\mathrm{sat}} /N_0 F$ for the four modulation formats of the standard (QPSK, 8PSK, 16APSK, and 32APSK).
We verified that comparable SE values can be also obtained by using, instead of the predistorter, the advanced detection scheme of Section~\ref{ss:advanced_detection} with $L_r=0$ (MMSE detection). We also consider two alternative ways that, at least in the case of a linear channel, can be used to improve the SE without resorting to TF packing.\footnote{On a nonlinear satellite channel, due to the increased peak-to-average power ratio, their application must be carefully considered since not necessarily produces the expected benefits.} The simplest approach relies on the increase of the modulation cardinality, and in Fig.~\ref{f:ASE_DVB} we also show the SE for the 64APSK modulation~\cite{LiAl08}.
It can be seen that the 64APSK modulation, due to the higher impact of the nonlinearities, allows to increase the SE only at high SNR values and it seems there is no hope to improve the SE in the low and medium SNR regions.

\begin{figure}
\centering{} \includegraphics[width=92mm]{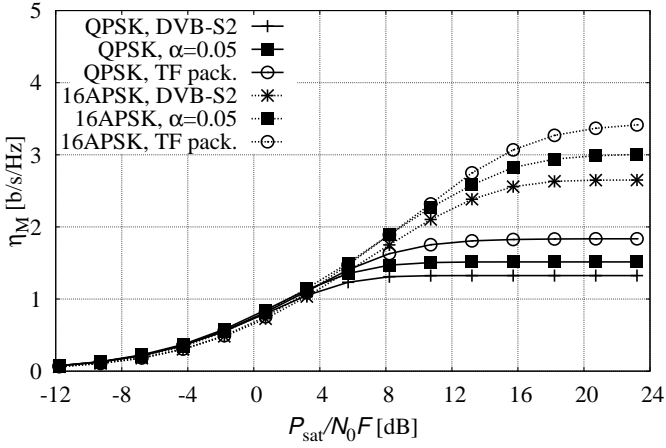}
 \caption{Improvements, in terms of spectral efficiency, that can be obtained for QPSK and 16APSK modulations by reducing of the roll-off ($\alpha=0.05$) or by adopting the TF packing technique. 
In all cases, predistortion at the transmitter and memoryless detection at the receiver are employed.}\label{f:ASE_TF_pre}
\end{figure}

An alternative way of improving the SE is based on a reduction of the roll-off factor. In Fig.~\ref{f:ASE_TF_pre}, we consider QPSK and 16APSK modulations in a scenario where predistortion at the transmitter and symbol-by-symbol detection at the receiver are still employed. We show the SE improvement that can be obtained by reducing the roll-off to $\alpha=0.05$.\footnote{We properly modify the transmitted signal such that it occupies the same bandwidth as that of the signal with roll-off 0.2. We verified that no improvement can be obtained by resorting to a more sophisticated receiver based on linear or nonlinear equalization in addition to or in substitution of the predistorter.}
We can observe that the roll-off reduction improves the SE with respect to DVB-S2 for all SNR values. On the other hand, as shown in Fig.~\ref{f:ASE_TF_pre}, better results can be obtained by allowing 
TF packing. The values of $T$ and $F$ are chosen as those providing the largest SE. This search is carried out by evaluating~(\ref{e:eta_M}) on grid of values of $T$ and $F$ (coarse search), followed by interpolation 
of the obtained values (fine search). We point out that these curves have been obtained without 
reducing the roll-off factor, which is still $\alpha=0.2$, and employing the same predistorter and symbol-by-symbol receiver adopted in the DVB-S2 system.

\begin{figure}
\centering{} \includegraphics[width=92mm]{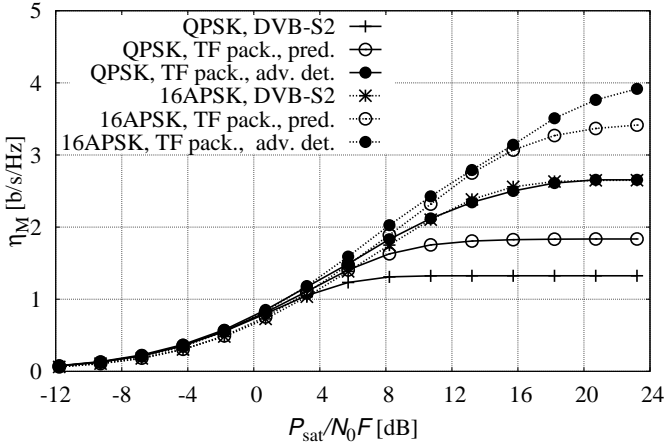}
 \caption{Spectral efficiency for QPSK and 16APSK modulations with TF packing and advanced detector (TF pack., adv. det.). Comparison with the DVB-S2 scenario and the case of TF packing when a predistorter and a symbol-by-symbol detector are adopted (TF pack., pred.).}\label{f:ASE_TF_CS}
\end{figure}

With the aim of further improving the performance, we now consider TF packing and a system without predistortion at the transmitter but using the advanced detection algorithm described in Section~\ref{ss:advanced_detection}, joint with CS  ($L_r=1$). 
The Volterra order of the model~(\ref{e:volterra_apprx}) is $v=5$. The results for QPSK and 16APSK modulations are reported in Fig.~\ref{f:ASE_TF_CS}, where we also show the DVB-S2 benchmark curves  discussed above and the curves related to TF packing 
when predistortion at the transmitter and memoryless detection at the receiver are used. These results show the impressive improvement achievable by TF packing combined with the considered advanced receiver, which, with a memory of only one symbol, can cope with much more interference than the schemes employing the predistorter and a memoryless detector.

\begin{figure}
 \centering{}\includegraphics[width=92mm]{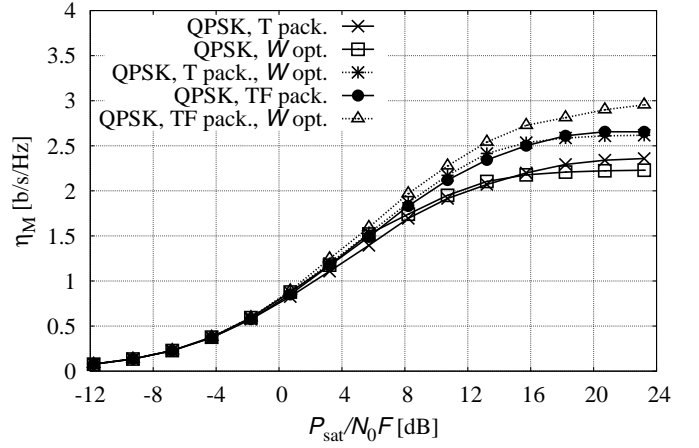}
 \caption{Spectral efficiency of QPSK modulation with TF packing and bandwidth optimization by adopting the advanced receiver with CS ($L_r=1$).}\label{f:ASE_B_PACK}
\end{figure}

Having removed the constraint of orthogonal signaling,
one more degree of freedom in the SE optimization is represented by the bandwidth $W$ of the shaping pulse $p(t)$ (in case of orthogonality, it is $W=(1+\alpha)/T$ instead).\footnote{The symbol rate, or equivalently the bandwidth, of conventional  DVB-S2 can also be further optimized for a performance improvement.} Hence, guided by the same idea behind the TF packing technique, we also optimized $W$, further increasing both ICI and ISI due to the adjacent users and to the IMUX and OMUX filters, respectively.
Whereas on the AWGN channel this optimization is implicit in TF packing, in the sense that we can obtain the same ICI by fixing $F$ and increasing $W$ or by fixing $W$ and decreasing $F$, 
this is no more true for our nonlinear channel since IMUX and OMUX bandwidths are kept fixed. Hence, an increased value of $W$ also increases the ISI.
The benefit of the bandwidth optimization is twofold: it can be used  as an alternative to frequency packing (e.g., in cases where the frequencies of the on-board oscillators cannot be modified and, hence, frequency packing is not an option), 
or it can be used to improve the results of TF packing.
In Fig.~\ref{f:ASE_B_PACK}, we consider QPSK modulation and the advanced receiver with $L_r=1$. As expected, the combination of TF packing with the bandwidth optimization gives the best results. We also show the results in case only time packing or only the bandwidth optimization are adopted. Interestingly, the SE of time packing with bandwidth optimization is quite similar to that achievable by TF packing.

\begin{figure}
\centering{} \includegraphics[width=92mm]{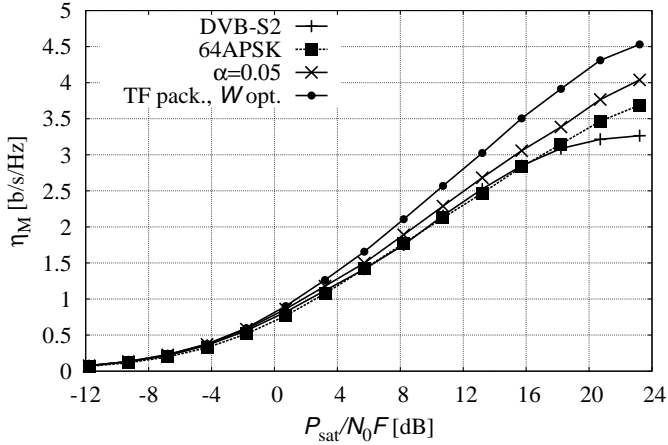}
 \caption{Spectral efficiency of TF packing with bandwidth optimization (TF pack., $W$ opt.). Comparison with DVB-S2, 64APSK and roll-off reduction.}\label{f:ASE_TFB}
\end{figure}

Finally, to summarize the results, Fig.~\ref{f:ASE_TFB} shows the SE for all DVB-S2 modulations (QPSK, 8PSK, 16APSK and 32APSK) with TF packing, bandwidth optimization, and the advanced receiver. For clarity, 
we show only one curve which, for each abscissa, reports only the largest value of the four curves (the ``envelope''). In the same figure, we also plot three other SE curves obtained by using predistortion and a memoryless receiver. The lowest one is that corresponding to the DVB-S2 scenario (one curve which is the ``envelope'' of all four curves in Fig.~\ref{f:ASE_DVB}), the SE curve for the 64APSK modulation, and the SE curve in  case of roll-off $\alpha=0.05$ reduction. In this latter case, we considered all modulations with cardinality up to 64, and hence this curve represents the effect 
of both roll-off reduction and cardinality increase with respect to DVB-S2. The figure shows that TF packing and advanced receiver processing allows a SE improvement of around 40$\%$ w.r.t. DVB-S2 at high SNR.

\subsection{Modulation and coding formats}
What information theory promises can be approached by using proper coding schemes. All the considered modulation and coding formats (MODCODs) use the low-density parity-check (LDPC) codes with length 64800 bits of the DVB-S2 
standard.  We adopt the optimized values for $T$, $F$, and $W$ and the advanced detector described in Section~\ref{ss:advanced_detection}.
We assume that $1/T_B=27.5$~Mbaud and $F_B=41.5$ MHz, and use these values to normalize the time and frequency spacings. 
Due to the soft-input soft-output nature of the considered detection algorithm, we can adopt iterative detection and decoding. We distinguish between local iterations, within the LDPC decoder, and global iterations,
between the detector and the decoder. Here, we allow a maximum of 5 global iterations and 20 local iterations.

\begin{figure}
\centering{} \includegraphics[width=92mm]{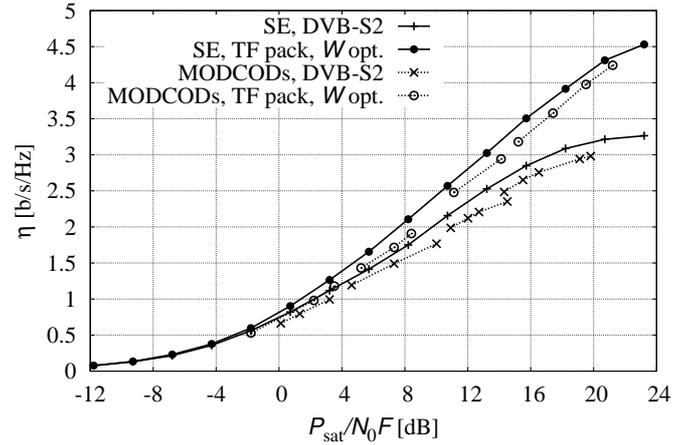}
 \caption{Modulation and coding formats of the DVB-S2 standard and comparison with those designed for the proposed TF packing technique with optimized bandwidth.}\label{f:modcods}
\end{figure}

BER results have been computed by means of Monte Carlo simulations and are reported in the
SE plane in Fig.~\ref{f:modcods} using, as reference, a BER of $10^{-6}$. In the same figure, the performance of the DVB-S2 MODCODs is also shown for comparison. 
We recall that for them predistortion at the transmitter and symbol-by-symbol detection at the receiver are adopted. Moreover, for them we have $\tau=1$ and $\nu=1$.
The details of the considered MODCODs are reported in Tables~\ref{t:modcods_det} and~\ref{t:modcods_dvbs2}.
These results are in perfect agreement with the theoretical analysis and confirm that the TF packing technique can provide an impressive performance improvement w.r.t. the DVB-S2 standard. 

\begin{table}
\caption{Details of the MODCODs based on TF packing.}\label{t:modcods_det}
    \centering
    \begin{tabular}{c|c|c|c|c|c|c|}
	\cline{2-7}
	& rate & $\tau$ & $\nu$ & $W_{\rm opt}$  &  $\!P/N_0 F\!$  & $\eta$ \\
	&   &   &   &    &    [dB] &   [b/s/Hz]\\
	\hline
	\hline
	\multirow{3}{*}{QPSK} & 1/3 & 0.833  & 1.00 & +20\% & -1.7 & 0.53 \\
	& 1/2 & 0.750 & 0.90 & +20\% & 2.2 & 0.98  \\
	& 3/5 & 0.750 & 0.90 & +20\% & 3.6 & 1.18 \\ 
	\hline
	\multirow{3}{*}{8PSK} & 1/2 & 0.731 & 0.95 & +30\%  & 5.3 & 1.43  \\
	& 3/5 & 0.731 & 0.95 & +30\%  & 7.4 & 1.72  \\
	& 2/3 & 0.731 & 0.95 & +30\% & 8.5 & 1.91  \\
	\hline
	\multirow{2}{*}{16APSK} & 2/3 & 0.792 & 0.90 & +20\% & 11.1 & 2.48  \\
	& 3/4 & 0.750 & 0.90& +20\%  & 14.1 & 2.94  \\
	\hline
	\multirow{4}{*}{32APSK} & 2/3 & 0.731 & 0.95 & +30\% & 15.3 & 3.18  \\
	& 3/4 & 0.731 & 0.95 & +30\%  & 17.5 & 3.58 \\ 
	& 5/6 & 0.731 & 0.95 & +30\%  & 19.5 & 3.98 \\ 
	& 8/9 & 0.731 & 0.95 & +30\%  & 21.2 & 4.24 \\ 
	\hline
    \end{tabular}
\end{table}

\begin{table}
   
     \caption{Details of the DVB-S2 MODCODs. In this case, $\tau=1$ and $\nu=1$.}\label{t:modcods_dvbs2}
     \centering
    \begin{tabular}{c|c|c|c|}
	\cline{2-4}
	& rate &  $P/N_0 F$ [dB] & $\eta$ [b/s/Hz]\\
	\hline
	\hline
	\multirow{3}{*}{QPSK} & 1/2 & 0.1 & 0.66 \\
	& 3/5  & 1.4 & 0.79  \\
	& 3/4 & 3.2 & 0.99 \\ 
	\hline
	\multirow{3}{*}{8PSK} & 3/5 & 4.6 & 1.19 \\
	& 3/4 & 7.3 & 1.49  \\
	& 8/9 & 10.0 & 1.77  \\
	\hline
	\multirow{2}{*}{16APSK} & 3/4 & 10.9 & 1.99  \\
	& 4/5 & 12.0 & 2.12  \\
	& 5/6 & 12.7 & 2.21  \\
	& 8/9 & 14.6 & 2.35  \\
	\hline
	\multirow{4}{*}{32APSK} & 3/4  & 14.3 & 2.48  \\
	& 4/5 & 15.6 & 2.65 \\ 
	& 5/6 & 16.5 & 2.76 \\ 
	& 8/9 & 19.2 & 2.94 \\ 
	& 9/10 & 19.9 & 2.98 \\ 
	\hline
    \end{tabular}
\end{table}

\section{Conclusions}\label{s:conclusions}
We have investigated the TF packing technique, joint with an advanced processing at the receiver, to improve the spectral efficiency of a nonlinear satellite system employing linear modulations with finite constellations.  As a first step, through an information-theoretic analysis, we computed the spectral efficiency achievable through this technique showing, with reference to the DVB-S2 scenario, that without an advanced processing at the receiver, the potential gains are very limited. On the other hand, a detector which takes into account a memory of only one symbol, and thus with a very limited complexity increase, it is possible to obtain a gain up to 40\% in terms of spectral efficiency with respect to the conventional use of the current standard. All these considerations can be extended to other channels and scenarios.

 \bibliography{Refs.bib,Refs_andre.bib}
\bibliographystyle{ieeetr}

\end{document}